\documentclass[aps,prb]{revtex4}
\usepackage{graphicx}

\begin{document}
\pagestyle{empty}
\title{Blackbody friction force on a relativistic small neutral particle}
\author{A. I. Volokitin$^{1,2}$\footnote{
\textit{E-mail address}:alevolokitin@yandex.ru}    }
 \affiliation{$^1$Peter Gr\"unberg Institut,
Forschungszentrum J\"ulich, D-52425, Germany}

\affiliation{
$^2$Samara State Technical University, Physical Department, 443100 Samara, Russia}
\affiliation{
$^3$Samara State Aerospace University, Physical Department, 443086 Samara, Russia}

\begin{abstract}

 The friction force acting on a small neutral particle during relativistic motion relative to the blackbody radiation is calculated in the framework of  fluctuation electrodynamics. It is shown that the particle acceleration is determined by the friction force in the particle rest reference frame ($K^{\prime}$-frame) which in general is not equal to the friction force in the  frame of the blackbody radiation ($K$-frame). The difference between the friction forces in the different frames is connected with   the change of  the rest mass  of a particle due to the absorption and emission of radiation. The friction force in the $K^{\prime}$-frame is determined  only by the interaction of a particle with the blackbody radiation.  In the $K$-frame the interaction of a particle with its own thermal radiation also contributes to the friction force. For the steady state temperature of a particle the friction forces in the $K^{\prime}$-and $K$-frames are equal. For an atom the blackbody friction is determined by the electronic linewidth broadening which is calculated considering the interaction of an atom with its own radiation. In the ultrarelativistic case ($1-\beta\rightarrow 0$) for an atom the friction force diverges as $(1-\beta)^{-3}$ and the (average) temperature of an atom $T_2\approx (1-\beta)^{-3/8}T_1$, where $T_1$ is the temperature of the blackbody radiation and $\beta=V/c$. Controversies in the theory of the blackbody friction are discussed.

\end{abstract}

\maketitle

\section{Introduction}

A remarkable example of the forces induced by fluctuations is the   friction force acting on a particle at motion relative to the  fluctuating electromagnetic field  created by thermal and quantum fluctuations. At present the radiation friction is attracting a lot of attention due to the fact that it
is one of the mechanisms of non-contact friction between bodies in the absence of direct contact \cite{VolokitinRMP2007}. The non-contact friction determines the ultimate limit to which the friction force can be reduced and, consequently, also the force fluctuations. The force
fluctuations (and hence friction) are important for ultrasensitive force registration. Perhaps the most
exciting application of these ideas is associated with mechanical detection of nuclear spin resonance  \cite{RugarNature2004}. For example, a
single spin detection using magnetic resonance force microscopy \cite{SidlesRMP1995} (which was proposed to obtain images of biological
objects, such as proteins, with atomic resolution) and for quantum computing \cite{BermanPRB200} would require reducing the fluctuating
forces (and therefore friction) to unprecedented levels. In addition, the measurements of  gravitation interactions at a small length scale \cite{ArkaniPLB1998}, and the future measurements of the Casimir forces \cite{SushkovNature2011} and the behavior of micro-and nano-electromechanical devices may eventually be limited by non-contact friction effects.

Radiative friction has deep roots in the foundation of quantum mechanics. Friction force during the motion of a particle relative to the blackbody radiation, which is a limiting case of the  fluctuating electromagnetic field,  was  studied by Einstein and Hopf \cite{Einstein1910}   and  Einstein \cite{Einstein1917} in the early days of quantum mechanics. The friction in this case can be explained by the direction-dependent Doppler effect. In the rest reference frame an atom absorbs blue-shifted counterpropagating blackbody photons, while emitting these photons in all directions which leads to the momentum transfer and friction.  Blackbody radiation friction is connected with the interaction of atoms with the cosmic blackbody radiation \cite{GreiserPRL1966,ZatsepinJETPLett1966,MkrtchianPRL2003}. The friction induced by  electromagnetic fluctuations was studied in the  plate-plate \cite{PendryJPCM1997,VolokitinJPCM1999,VolokitinPRL2003,VolokitinPRB2003,VolokitinRMP2007,VolokitinPRB2008},  neutral particle-plate \cite{TomassonePRB1997,VolokitinPRB2002,VolokitinRMP2007,DedkovPLA2005,DedkovJPCM2008,BrevikEntropy2013,KardarPRD2013,BrevikEPJD2014,BartonNJP2010,HenkelNJP2013,
DalvitPRA2014}, and neutral particle-blackbody radiation  \cite{VolokitinRMP2007,VolokitinPRB2008,HenkelNJP2013,MkrtchianPRL2003,DedkovPLA2005,DedkovNIMPR2010,DedkovPhysScr2014,JentschuraPRL2012,JentschuraPRL2015} configurations. However, the theory of radiative friction is still controversial. As an example, different authors have studied the  friction force between a small neutral particle and blackbody radiation \cite{VolokitinRMP2007,VolokitinPRB2008,HenkelNJP2013,MkrtchianPRL2003,DedkovPLA2005,DedkovNIMPR2010,DedkovPhysScr2014,JentschuraPRL2012,JentschuraPRL2015}, using different approaches, and have obtained results which are in sharp contradiction
to each other. In Refs. \cite{Einstein1910,Einstein1917,VolokitinRMP2007,VolokitinPRB2008,MkrtchianPRL2003}  only interaction of a particle with blackbody radiation was taken into account, therefore the friction force depends only on the temperature of this radiation. On the other hand, in Refs. \cite{HenkelNJP2013,DedkovPLA2005,DedkovNIMPR2010} the interaction of a particle with its own thermal radiation was also taken into account, in this case the friction force depends on  the temperatures of both a particle and the blackbody radiation. Controversies between different theories of blackbody radiation friction recently were discussed in  Ref.  \cite{DedkovNIMPR2010}.

In the present paper  a general theory of the blackbody friction for a neutral particle at relativistic motion relative to the blackbody radiation which includes as limiting cases the previous theories from Refs.\cite{VolokitinRMP2007,VolokitinPRB2008,HenkelNJP2013,MkrtchianPRL2003,DedkovPLA2005,DedkovNIMPR2010,DedkovPhysScr2014,JentschuraPRL2012,JentschuraPRL2015} is developed. This general theory establishes a link between the different theories of the blackbody friction.  In sharp contrast to the opinion of the authors of Ref.\cite{DedkovNIMPR2010}  the total friction forces acting on a  particle   in the frame of the blackbody radiation ($K$-frame) and in the rest frame of a particle ($K^{\prime}$-frame) are not equal.  This difference is due to the change of the rest mass of a particle as a result of the absorption and emission of radiation by a particle. Recently, Ref. \cite{HenkelNJP2013} used a fully covariant formulation of the blackbody friction. However, the physical origin of the difference of the friction force in the different frame and the condition for the steady state temperature of a particle were not established in that article.

\section{Theory}

 We introduce two reference frames, $K$ and $K^{\prime}$.
 $K$ is the frame of blackbody radiation, and $K^{\prime}$ moves with  velocity
 $V$ along the $x$ axis.
In the $K^{\prime}$  frame  a particle moves with  velocity
 $v^{\prime}\ll V$ along the $x$-axis. At the motion of a particle with acceleration the $K^{\prime}$ frame coincides  with the particle rest reference frame only at $t=t_0$ when  $v^{\prime}(t_0)=0$. However, for  $v^{\prime}\ll V$ the difference between the friction forces in the $K^{\prime}$ frame and the particle rest reference frame can be neglected thus in this paper the $K^{\prime}$ frame is denoted as the particle rest reference frame. The relation between the $x$ components of the momentum in the different reference frames is given by
\begin{equation}
p_x=(p_x^{\prime}+\beta E^{\prime}/c)\gamma, \label{1bb1}
\end{equation}
where $\beta=V/c$, $\gamma = 1/\sqrt{1-\beta^2}$, $E^{\prime}=E_0/\sqrt{1-(v^{\prime}/c)^2}$ is the total energy of a particle in the $K^{\prime}$ frame, and $E_0=m_0c^2$  is the rest energy of a particle. The rest energy can change due to the absorption and emission of thermal radiation by a particle.
The connection between forces in the    $K$ and $K^{\prime}$ frames follows from Eq. (\ref{1bb1})
\begin{equation}
\frac{dp_x}{dt}=\frac{1}{1+(Vv^{\prime}/c^2)}\left[\frac{dp_x^{\prime}}{dt^{\prime}}+V
\frac{dm_0}{dt^{\prime}}\frac{1}{\sqrt{1-(v^{\prime}/c)^2}}+m_0V\frac{d}{dt^{\prime}}\left(\frac{1}{\sqrt{1-(v^{\prime}/c)^2}}\right)\right]. \label{2bb1}
\end{equation}
For $v^{\prime}\ll V$ from (\ref{2bb1}) we get
\begin{equation}
F_x=F_x^{\prime}+V
\frac{dm_0}{dt^{\prime}},
 \label{2bb11}
\end{equation}
where $F_x$ and $F_x^{\prime}$ are the forces in the   $K$ and $K^{\prime}$ frames, respectively.
The last term in Eq. (\ref{2bb11}) determines the rate of change of the momentum of a particle in the   $K$ frame due to the change of its rest mass as a result of the absorption and emission of radiation by a particle.
Taking into account that at  $v^{\prime}\ll V$
\begin{equation}
\frac{dp_x}{dt} =
\frac{d}{dt}\left(\frac{m_0v}{\sqrt{1-(v/c)^2}}\right)=\frac{dm_0}{dt^{\prime}}V +
m_0\gamma^3\frac{dv}{dt}, \label{4bb1}
\end{equation}
from Eq. (\ref{2bb11}) we get
\begin{equation}
m_0\gamma^3\frac{dv}{dt}=\frac{dp_x^{\prime}}{dt^{\prime}}=F_x^{\prime},
\label{4bc}
\end{equation}
where $v$ is the velocity of a particle in the $K$ frame.
From Eq. (\ref{4bc}) it follows that acceleration in the  $K$ frame is determined by the friction force in the   $K^{\prime}$ frame.

At uniform motion of a particle an external force $f_x$ should be applied to it. At $V=$const the equation of the motion of a particle is
\begin{equation}
\gamma V\frac{dm_0}{dt}=F_x+f_x.
\label{equation1}
\end{equation}
If the force  $f_x$ does not change the rest mass of a particle, then its value is the same in the   $K$ and $K^{\prime}$ frames, i.e.,  $f_x=f_x^{\prime}$. In this case
\begin{equation}
f_x=-F_x+\gamma V\frac{dm_0}{dt}=-F_x+\frac{dm_0}{dt^{\prime}}V=-F_x^{\prime},
\label{equation2}
\end{equation}
that is,  for  uniform motion of a particle, an external force that is  equal but  opposite in sign to the friction force in the  $K^{\prime}$ frame should be applied to it.

In the $K^{\prime}$ frame the Lorentz force
on the particle is determined by the expression \cite{VolokitinPRB2002,Dedkov2008}

\begin{equation}
F_x^{\prime} = \frac{\partial}{\partial x ^{\prime}}\langle
\mathbf{d_e^{\prime}\cdot E^{\prime
*}(r^{\prime})}\rangle_{\mathbf{r^{\prime} =r^{\prime}_0}} , \label{6bb}
\end{equation}
where according to the fluctuation electrodynamics  \cite{Rytov53,Rytov67,Rytov89}
$\mathbf{d}_e^{\prime}=\mathbf{d}_e^{f\prime}+\mathbf{d}_e^{in\prime}$, $\mathbf{E}^{\prime}=
\mathbf{E}^{f\prime}+\mathbf{E}^{in\prime}$, $\mathbf{d}_e^{f\prime}$ and $\mathbf{E}^{f\prime}$
are the fluctuating dipole moment of a particle and the electric field of the blackbody radiation,
and $\mathbf{d}_e^{in\prime}$ and $\mathbf{E}^{in\prime}$ are the dipole moment of a particle
induced by the blackbody radiation and the electric field induced by the fluctuating dipole moment
of a particle, respectively. Taking into account the statistical independence of the fluctuating values, the Lorentz force can be written in the form
\begin{equation}
F_x^{\prime} = F_{1x}^{\prime}+ F_{2x}^{\prime}, \label{6bb2}
\end{equation}
where
\begin{equation}
F_{1x}^{\prime} =\frac{\partial}{\partial x ^{\prime}}\langle
\mathbf{d_e^{in\prime}\cdot E^{f\prime
*}(r^{\prime})}\rangle_{\mathbf{r^{\prime} =r^{\prime}_0}}, \label{6bb2}
\end{equation}
\begin{equation}
F_{2x}^{\prime} =\frac{\partial}{\partial x ^{\prime}}\langle
\mathbf{d_e^{f\prime}\cdot E^{in\prime
*}(r^{\prime})}\rangle_{\mathbf{r^{\prime} =r^{\prime}_0}}. \label{6bb3}
\end{equation}

To calculate  $F_{1x}^{\prime}$ while writing the electric field in the $K^{\prime}$ frame as a Fourier integral
\[
\mathbf{E}^{f\prime
}(\mathbf{r}^{\prime}, t^{\prime})
= \int_{-\infty}^{\infty}
\frac{d\omega^{\prime}}{2\pi}\int
\frac{d^3k^{\prime}}{(2\pi)^3}e^{i\mathbf{k}^{\prime}\cdot\mathbf{r}^{\prime}-i\omega^{\prime}t^{\prime}}\mathbf{E}^{f\prime
}(\mathbf{k}^{\prime}, \omega^{\prime}),
\]
and taking into account that
\[
\mathbf{d}_e^{in\prime} =
\int_{-\infty}^{\infty}
\frac{d\omega^{\prime}}{2\pi}\int
\frac{d^3k^{\prime}}{(2\pi)^3}\alpha(\omega^{\prime})e^{i\mathbf{k}^{\prime}\cdot\mathbf{r}^{\prime}-i\omega^{\prime}t^{\prime}}\mathbf{E}^{f\prime
}(\mathbf{k}^{\prime}, \omega^{\prime}),
\]
where $\alpha(\omega^{\prime})$  is the particle polarizability,
we get
\begin{equation}
F_{1x}^{\prime}
=-i\int_{\infty}^{\infty} \frac{d\omega^{\prime}}{2\pi}\int
\frac{d^3k^{\prime}}{(2\pi)^3}k_x^{\prime}
\alpha(\omega^{\prime})\langle \mathbf{E^{f\prime}\cdot
E^{f\prime *}}\rangle_{\omega^{\prime}\mathbf{k}^{\prime}}.
\label{7bb}
\end{equation}
When we change
from the $K^{\prime}$ frame to the
 $K$ frame $\langle \mathbf{E^{\prime}\cdot E^{\prime *}}\rangle_{\omega^{\prime}\mathbf{k}^{\prime}}$ is transformed as the energy density of
a plane electromagnetic field. From the law of transformation of the energy density of
a plane electromagnetic field \cite{LandauField} we get
\begin{equation}
\langle \mathbf{E^{f\prime}\cdot E^{f\prime
*}}\rangle_{\omega^{\prime}\mathbf{k}^{\prime}} = \langle
\mathbf{E^f\cdot E^{f *}}\rangle_{\omega
\mathbf{k}}\left(\frac{\omega^{\prime}}{\omega}\right)^2.
\label{8bb}
\end{equation}
According to the theory of the fluctuating electromagnetic field the spectral density of the fluctuations of the electric field is determined by
\cite{Lifshitz80}
\begin{equation}
<E_i^f(\mathbf{r})E_j^{f*}(\mathbf{r^{\prime}})>_{\omega \mathbf{k}}
=\hbar\mathrm{Im}D_{ij}(\mathbf{k},
\omega)\coth\left(\frac{\hbar
\omega}{2k_BT_1}\right),\label{thelh3}
\end{equation}
where the Green's function of the electromagnetic field in the free space is determined by
\begin{equation}
D_{ik}(\omega, \mathbf{k}) = -
\frac{4\pi\omega^2/c^2}{\omega^2/c^2 - k^2 + i0\cdot \mathrm{sgn}
\,\omega }\left[\delta_{ik} - \frac{c^2k_ik_k}{\omega^2
}\right], \label{thelh2a}
\end{equation}
 $T_1$ is the temperature of the blackbody radiation. Taking into account that
\[
\mathrm{Im}\frac{1}{\omega^2/c^2 - k^2 + i0\cdot \mathrm{sign}
\,\omega }=
\mathrm{Im}\frac{1}{\omega^{\prime 2}/c^2 - k^{\prime 2} + i0\cdot \mathrm{sgn}
\,\omega^{\prime } },
\]
we get
\begin{equation}
\langle \mathbf{E^{f\prime}\cdot E^{f\prime*}}\rangle_{\omega^{\prime} \mathbf{k}^{\prime}} = 4\pi^2
\hbar k^{\prime} \left\{\delta (\frac{\omega^{\prime}}{c}-k^{\prime}) - \delta
(\frac{\omega^{\prime}}{c}+k^{\prime})\right\}\coth\left(\frac{\hbar
\omega}{2k_BT_1}\right). \label{9bb}
\end{equation}

Substitution (\ref{9bb}) in Eq. (\ref{7bb}) and integration over
 $\omega^{\prime}$ give
\begin{equation}
F_{1x}^{\prime} = \frac{\hbar c}{2\pi^2}\int d^3k^{\prime} k^{\prime} k_x^{\prime}
\mathrm{Im}\alpha (ck^{\prime})
\coth\left(\frac{\hbar\gamma
(ck^{\prime}+Vk_x^{\prime})}{2k_BT_1}\right),
\label{12bb}
\end{equation}
where it was taken into account, that  $\omega = (\omega^{\prime}
+ k^{\prime}_xV)\gamma$. Introducing the new variable  $\omega^{\prime} = ck^{\prime}$,
(\ref{12bb}) can be written in the form
\begin{equation}
F_{1x}^{\prime} = \frac{\hbar }{\pi c^2}\int_0^{\infty} d\omega^{\prime} \omega^{\prime2}
\int_{-\omega^{\prime}/c}^{\omega^{\prime}/c} dk_x^{\prime} k_x^{\prime} \mathrm{Im}\alpha (\omega^{\prime})
  \coth\left(\frac{\hbar\gamma
(\omega^{\prime}+Vk_x^{\prime})}{2k_BT_1}\right).
\label{13bb}
\end{equation}
At small velocities ($V\ll c$)$ F_x = -\Gamma V$, where
\begin{equation}
\Gamma =\frac{ \hbar ^2}{3\pi c^5k_BT_1}\int_0^\infty d\omega
\frac{ \omega ^5\mathrm{Im}\alpha (\omega )}{\sinh ^2(\frac{
\hbar \omega}{2k_BT_1}  )}, \label{bb1}
\end{equation}
Equation (\ref{bb1}) was first derived in Ref. \cite {MkrtchianPRL2003} using a different approach.
The rate of change of the rest energy of a particle in the  $K^{\prime}$ frame due to the absorption of blackbody radiation is determined by the equation
\begin{equation}
P_1^{\prime}=\frac{dm_0}{dt^{\prime}}c^2=\langle
\mathbf{j}_e^{in\prime}\cdot \mathbf{E}^{f\prime
*}\rangle =
 \frac{\partial}{\partial t ^{\prime}}\langle
\mathbf{d}_e^{in\prime}(t^{\prime})\cdot \mathbf{E}^{f\prime
*}(t^{\prime}_0)\rangle_{t^{\prime} =t^{\prime}_0} . \label{6bb}
\end{equation}
After the calculations, which are similar to what was done when calculating   $F_{1x}^{\prime}$ we get
\begin{equation}
P_1^{\prime} = \frac{\hbar }{\pi c^2}\int_0^{\infty} d\omega^{\prime} \omega^{\prime2}
\int_{-\omega^{\prime}/c}^{\omega^{\prime}/c} dk_x^{\prime} \omega^{\prime} \mathrm{Im}\alpha (\omega^{\prime})
 \coth\left(\frac{\hbar
\gamma(\omega^{\prime}+Vk_x^{\prime})}{2k_BT_1}\right).
\label{13bbh}
\end{equation}
From Eq. (\ref{2bb11})  the friction force acting on a particle in the  $K$-frame due to the interaction with the blackbody radiation is given by
\[
F_{1x}=F_{1x}^{\prime}+\beta \frac{P_1^{\prime}}{c}
\]
\begin{equation}
 = \frac{\hbar }{\pi c^2}\int_0^{\infty} d\omega^{\prime} \omega^{\prime2}
\int_{-\omega^{\prime}/c}^{\omega^{\prime}/c} dk_x^{\prime} (k_x^{\prime}+\beta\frac{\omega^{\prime}}{c}) \mathrm{Im}\alpha (\omega^{\prime})
 \coth\left(\frac{\hbar
\gamma(\omega^{\prime}+Vk_x^{\prime})}{2k_BT_1}\right).
\label{14bb}
\end{equation}
Introducing  the new variables: $k_x^{\prime}=\gamma(q_x-\beta\omega/c)$, $\omega^{\prime}=\gamma(\omega-Vk_x)$ in the integral (\ref{14bb}) we get
\begin{equation}
F_{1x} = \frac{\hbar\gamma }{\pi c^2}\int_0^{\infty} d\omega
\int_{-\omega/c}^{\omega/c} dk_x k_x (\omega-Vk_x)^2\mathrm{Im}\alpha [\gamma(\omega-Vk_x)]
 \coth\left(\frac{\hbar
\omega}{2k_BT_1}\right),
\label{15bb}
\end{equation}
where we have taken into account that $d\omega^{\prime}dq_x^{\prime}=d\omega dq_x$.
To calculate $F_{2x}^{\prime}$ in the $K^{\prime}$ frame  we use the representation of the fluctuating dipole moment of a particle as a Fourier integral
\begin{equation}
\mathbf{d}^{f}( t^{\prime})
= \int_{-\infty}^{\infty}
\frac{d\omega^{\prime}}{2\pi}e^{-i\omega^{\prime}t^{\prime}}\mathbf{d}^{f
}( \omega^{\prime}).
\label{16bb}
\end{equation}
The electric field created in the  $K^{\prime}$ frame by the fluctuating dipole moment  of a particle is determined by the equation
\begin{equation}
E_i^{in\prime
}(\mathbf{r}^{\prime}, t^{\prime})
= \int_{-\infty}^{\infty}
\frac{d\omega^{\prime}}{2\pi}\int
\frac{d^3k^{\prime}}{(2\pi)^3}e^{i\mathbf{k}^{\prime}\cdot(\mathbf{r}^{\prime}-\mathbf{r}_0^{\prime})-i\omega^{\prime}t^{\prime}}
D_{ik}(\omega^{\prime}, \mathbf{k}^{\prime})d_k^{f
}( \omega^{\prime}).
\label{17bb}
\end{equation}
According to the fluctuation-dissipation theorem, the spectral density of the fluctuations of the fluctuating dipole moment is determined by the equation
\cite{Lifshitz80}
\begin{equation}
<d^f_id_k^{f*}>_{\omega^{\prime}}
=\hbar\mathrm{Im}\alpha(
\omega^{\prime})\coth\left(\frac{\hbar
\omega^{\prime}}{2k_BT_2}\right)\delta_{ik},\label{18bb}
\end{equation}
where  $T_2$ is the temperature of a particle.
Substituting Eqs. (\ref{16bb}) and  (\ref{17bb}) in Eq. (\ref{6bb3}) and taking into account Eq. (\ref{18bb}), we get
\begin{equation}
F_{2x}^{\prime} = -\frac{\hbar }{\pi c^2}\int_0^{\infty} d\omega^{\prime} \omega^{\prime2}
\int_{-\omega^{\prime}/c}^{\omega^{\prime}/c} dk_x^{\prime} k_x^{\prime} \mathrm{Im}\alpha (\omega^{\prime})
  \coth\left(\frac{\hbar
\gamma\omega^{\prime}}{2k_BT_2}\right)=0.
\label{19bb}
\end{equation}
Thus, in the   rest reference frame of a particle the friction force due to its own thermal radiation  is zero. This result
due to the fact that in this frame, due to the symmetry,  the total radiated
momentum  from  the  dipole  radiation  is
identically zero.  Thus the change in momentum of a particle  in
the rest reference frame is determined by the Lorentz force
$F_x^{\prime}$ acting on a particle from the external
electromagnetic field associated with the  blackbody radiation
observed in this reference frame.
The rate of change of the rest energy of a particle in the  $K^{\prime}$ frame due to its thermal radiation can be obtained with  similar calculations
\[
P_2^{\prime}=\langle
\mathbf{j}_e^{f\prime}\cdot \mathbf{E}^{in\prime
*}\rangle =
 \frac{\partial}{\partial t ^{\prime}}\langle
\mathbf{d}_e^{f\prime}(t^{\prime})\cdot \mathbf{E}^{in\prime
*}(t^{\prime}_0)\rangle_{t^{\prime} =t^{\prime}_0}
\]
\begin{equation}
 = -\frac{\hbar }{\pi c^2}\int_0^{\infty} d\omega^{\prime} \omega^{\prime2}
\int_{-\omega^{\prime}/c}^{\omega^{\prime}/c} dk_x^{\prime} \omega^{\prime} \mathrm{Im}\alpha (\omega^{\prime})
 \coth\left(\frac{\hbar
\gamma\omega^{\prime}}{2k_BT_2}\right),
\label{20bb}
\end{equation}
and the friction force in the $K$ frame associated with thermal radiation of a particle is given by
\begin{equation}
F_{2x} = F_{2x}^{\prime} + \beta\frac{P_2^{\prime}}{c}=-\frac{\hbar\gamma }{\pi c^2}\int_0^{\infty} d\omega
\int_{-\omega/c}^{\omega/c} dk_x k_x (\omega-Vk_x)^2\mathrm{Im}\alpha [\gamma(\omega-Vk_x)]
 \coth\left(\frac{\hbar
\gamma(\omega-Vk_x)}{2k_BT_2}\right),
\label{15bb}
\end{equation}
The total friction force in the $K$ frame is given by
\begin{equation}
F_x=F_{1x}+F_{2x} = \frac{2\hbar\gamma }{\pi c^2}\int_0^{\infty} d\omega
\int_{-\omega/c}^{\omega/c} dk_x k_x (\omega-Vk_x)^2\mathrm{Im}\alpha (\gamma(\omega-Vk_x))(n_1(\omega)-n_2(\omega^{\prime})),
\label{16bbf}
\end{equation}
where $n_i(\omega)=[\exp(\hbar\omega/k_BT_i)-1]^{-1}$. Equation (\ref{16bbf}) was first derived in Ref. \cite{DedkovPLA2005}. Recently, it was derived on the basis of a relativistically covariant theory \cite{HenkelNJP2013}. The approach presented in this paper is more compact and transparent, and it gives us the possibility to connect the difference between the friction forces in the different frames to the absorption and emission of radiation by a particle.   Note that the friction force $F_x$ can be either positive or negative. However, the acceleration, which is determined by the friction force
$F_x^{\prime}$, is always negative.
The total heat absorbed by a particle in the  $K^{\prime}$-frame is determined by the equation
\begin{equation}
P^{\prime}=P^{\prime}_1+P^{\prime}_2=\frac{2\hbar\gamma^2 }{\pi c^2}\int_0^{\infty} d\omega^{\prime}
\int_{-\omega^{\prime}/c}^{\omega^{\prime}/c} dk_x^{\prime}\omega ^{\prime} (\omega-Vq_x)^2\mathrm{Im}\alpha [\gamma(\omega-Vq_x)](n_1(\omega)-n_2(\omega^{\prime})).
\label{17bbh}
\end{equation}
In the  $K$ frame the total change in energy of a particle due to the interaction with radiation  can be calculated from the law of the transformation of energy of a particle: $E=\gamma(E^{\prime}+p^{\prime}_xV)$, where $E$ and $E^{\prime}$ are the total energy of a particle in the   $K$ and $K^{\prime}$ frames, respectively. From this relation we get the equation for the rate of change of the energy of a particle in the $K$ frame
\begin{equation}
\frac{dE}{dt}=P=P^{\prime}+F_x^{\prime}V=\frac{2\hbar\gamma }{\pi c^2}\int_0^{\infty} d\omega
\int_{-\omega/c}^{\omega/c} dk_x\omega  (\omega-Vk_x)^2\mathrm{Im}\alpha [\gamma(\omega-Vk_x)][n_1(\omega)-n_2(\omega^{\prime})].
\label{18bbh}
\end{equation}
Formula  (\ref{17bbh}) was also recently obtained  in  \cite{DedkovPhysScr2014} using a different approach.
The rate of change of the energy of the blackbody radiation in the $K$ frame is determined by the equation $dW_{BB}/dt=-P$. The steady-state temperature of a particle is determined by the condition $P^{\prime}(T_1,T_2)=0$, and for this state $F_x=F_x^{\prime}$ and $P=F_x^{\prime}V$.

\section{Results}

The friction force acting on a particle moving relative to the blackbody radiation is determined by the imaginary part of the particle polarizability. For an atom the imaginary part of the polarizability is determined by the atom electronic linewidth broadening due to the radiation mechanism which can be calculated considering  the interaction of an atom with its own radiation.  Taking into account this interaction, the  dipole moment of an atom induced by an external electric field $E_x^{ext}(\omega,\mathbf{r}_0)$ can be written in the form \cite{VolokitinPRB2002,VolokitinPRB2006}
\begin{equation}
p_x^{ind}=\alpha_0(\omega)(\omega)[E_x^{ind}(\omega,\mathbf{r}_0)+E_x^{ext}(\omega,\mathbf{r}_0)],
\label{dipol1}
\end{equation}
where in the single-oscillator model without the radiation linewidth broadening the atomic polarizability  is given by  the equation
\begin{equation}
\alpha_0(\omega)=\frac{\alpha(0)\omega_0^2}{\omega_0^2-\omega^2},
\label{dipol2}
\end{equation}
$\alpha(0)$ is the static polarizability of an atom, and $E_x^{ind}(\omega,\mathbf{r}_0)$ is the radiation electric field created by the induced dipole moment of an atom. In the Coulomb gauge, which is used in this article, the Green's function of the electromagnetic field determines the electric field created by the unit point dipole, so $E_x^{ind}(\omega,\mathbf{r}_0)= \tilde{D}_{xx}(\omega,\mathbf{r}_0,\mathbf{r}_0)p_x^{ind}$, where $\tilde{D}_{xx}(\omega,\mathbf{r}_0,\mathbf{r}_0)$ is the reduced part of the Green's function of the electromagnetic field in the vacuum, which takes into account only the contribution from the propagating electromagnetic waves and determines the radiation  in the far field. The Green's function of the electromagnetic field in the vacuum $D_{xx}(\omega,\mathbf{r}_0,\mathbf{r}_0)$  diverges at  $\mathbf{r}=\mathbf{r}_0$. However, the contribution from the propagating waves remains finite and purely imaginary at  $\mathbf{r}=\mathbf{r}_0$, and the divergent contribution from the evanescent waves is real.  Therefore, $\tilde{D}_{xx}(\omega,\mathbf{r}_0,\mathbf{r}_0)=i\mathrm{Im}D_{xx}(\mathbf{r_0},\mathbf{r_0})$. From Eqs. (\ref{dipol1}) and (\ref{dipol2}) we get
\begin{equation}
\mathrm{Im}\alpha(\omega) = \mathrm{Im}\frac{p_x^{ind}}{E_x^{ext}(\omega,\mathbf{r}_0)}= \mathrm{Im}\frac{\alpha(0)\omega_0^2}{\omega_0^2-\omega^2-i\alpha(0)\omega_0^2\mathrm{Im}D_{xx}(\mathbf{r_0},\mathbf{r_0})}=
\frac{\alpha^2(0)\omega_0^4\mathrm{Im}D_{xx}(\mathbf{r_0},\mathbf{r_0})}{(\omega_0^2-\omega^2)^2+[\alpha(0)\omega_0^2\mathrm{Im}D_{xx}]^2},
\label{19bb}
\end{equation}
where
\begin{equation}
\mathrm{Im}D_{xx}(\mathbf{r_0},\mathbf{r_0})=\mathrm{Im}D_{yy}=\mathrm{Im}D_{zz}
= \int
\frac{d^3k}{(2\pi)^3}
\mathrm{Im}D_{xx}(\omega, \mathbf{k})=\frac{2}{3}\left(\frac{\omega}{c}\right)^3\mathrm{sgn}\,\omega.
\label{18bbd}
\end{equation}
At the resonance  ($\omega^2\approx\omega_0^2$) usually $\alpha(0)\mathrm{Im}D_{xx}\ll 1$ (for  example, for a hydrogen atom it is $\sim 10^{-6}$) thus the limit $\alpha(0)D_{xx}\rightarrow i0$ can be taken. In this case the resonant contribution is given by
\begin{equation}
\mathrm{Im}\alpha(\omega)\approx \frac{\pi\alpha(0)\omega_0}{2}[\delta(\omega-\omega_0)-\delta(\omega+\omega_0)],
\label{20bbd}
\end{equation}
and the off-resonant contribution  in the field far from resonance ($\omega^2\ll\omega_0^2$)
\begin{equation}
\mathrm{Im}\alpha(\omega)\approx \frac{2}{3}\left(\frac{\omega}{c}\right)^3\alpha^2(0)\mathrm{sign}\,\omega.
\label{21bbd}
\end{equation}
Recently, the result (\ref{21bbd}) was also obtained using quantum electrodynamics \cite{JentschuraPRL2015}. However, the analysis presented in this paper, is much simpler, and it clarifies the physical meaning of the terms of the series of the perturbation theory of quantum electrodynamics.
Using Eqs. (\ref{20bbd})  and (\ref{21bbd}) in Eq. (\ref{13bb}) we get the resonant and off-resonant contributions to the friction force
\begin{equation}
F_{1x}^{res}=\frac{ \hbar \omega^5_0\alpha(0)}{ c^4}\int_{-1}^1dxx\left[\exp{\left(\frac{\hbar\gamma\omega_0(1+\beta x)}{k_BT_1}\right)}-1\right]^{-1},
\label{21bbfr}
\end{equation}
\begin{equation}
F_{1x}^{nonres}=-\frac{ 512\pi^7\hbar \alpha(0)^2\gamma^6}{945 c^7 }\left(\frac{k_BT_1}{\hbar}\right)^8(7\beta+14\beta^3+3\beta^5),
\label{22bbfnr}
\end{equation}

At $\beta \ll 1$ the friction force $F_{1x}=-\Gamma V$, where   the resonant and off-resonant contributions to the friction coefficient are
\begin{equation}
\Gamma_{res}=\frac{ \hbar ^2\alpha(0)\omega^6_0}{6 c^5k_BT_1}
\frac{ 1}{\sinh ^2(\frac{
\hbar \omega_0}{2k_BT_1}  )},
\label{22bb}
\end{equation}
\begin{equation}
\Gamma_{nonres}=\frac{ 512\pi^7\hbar \alpha(0)^2}{135 c^8 }\left(\frac{k_BT_1}{\hbar}\right)^8,
\label{23bb}
\end{equation}
Results (\ref{22bb}) and (\ref{23bb}) were already obtained in Refs. \cite{MkrtchianPRL2003,JentschuraPRL2015}. In the ultrarelativistic case ($1-\beta \ll k_BT_1/\hbar\omega_0\ll 1$)
\begin{equation}
F_{1x}^{res}=\frac{  \omega^4_0\alpha(0)}{ c^4}\sqrt{2}k_BT_1\sqrt{1-\beta}\ln\frac{\hbar \omega_0\sqrt{1-\beta}}{\sqrt{2}k_BT_1},
\label{21bbfrr}
\end{equation}
\begin{equation}
F_{1x}^{nonres}=\frac{ 216\pi^7\hbar \alpha(0)^2}{135 c^7 (1-\beta)^3}\left(\frac{k_BT_1}{\hbar}\right)^8
\label{22bbfnrr}
\end{equation}
Thus in the ultrarelativistic case $F_{1x}^{res}\sim\sqrt{1-\beta}\ln\sqrt{1-\beta}\rightarrow 0$ and $F_{1x}^{nonres}\sim (1-\beta)^{-3}\rightarrow \infty$ at $1-\beta \rightarrow 0$.

For small velocities and typical temperatures the infrared thermal peak of the blackbody radiation is far below the resonance frequency  $\omega_0$, so it dominates the far-off-resonant contribution (\ref{22bb}), which has already been mentioned in Ref. \cite{JentschuraPRL2012}. However, in the ultrarelativistic case the friction is dominated by the far-off-resonant contribution for all temperatures.

According to Eq. (\ref{17bbh}) the total heat absorbed by an atom in the  $K^{\prime}$ frame is determined by the equation
\begin{equation}
P^{\prime}=\frac{ 128\pi^7 k_B^8\alpha(0)^2}{315 c^6\hbar^7 }\left[\gamma^6(7+35\beta^2+21\beta^4+\beta^6)T_1^8-7T_2^8\right].
\label{17bbhr}
\end{equation}
For small velocities the steady-state temperature of a particle $T_2=T_1$ and in the ultrarelativistic case $T_2\approx (1-\beta)^{-3/8}T_1$.
The problem in calculating  the imaginary part of the atom's  polarizability in the far off-resonant field  was considered in Ref. \cite{JentschuraPRL2012}. It was noted that in the literature there are still  questions still remain regarding the gauge invariance of the imaginary part of the polarizability. In this paper, the imaginary part of the atomic polarizability in the   field far from resonance is determined by the imaginary part of  the electric field of the unit point dipole, which is a gauge-invariant quantity. In the Coulomb gauge, which is used in this article, the electric field of the unit point dipole is the same as the Green's function of the electromagnetic field. At another gauge the expression for the Green's function will changed; however, the electric field determined with this Green's function will remain unchanged. Therefore, despite the fact that the  Green's function for the electromagnetic field is a gauge-dependent quantity, the imaginary part of the atomic polarizability calculated in this article is a gauge-invariant quantity. The gauge invariance of  the obtained results are also confirmed by the direct calculation using  quantum electrodynamics \cite{JentschuraPRL2015}. The gauge-invariant formulation  presented in this paper   confirms that the polarizability of the atom, for small frequencies, is a nonresonant effect, which is proportional to $\omega^3$ for small driving frequency $\omega$. This is consistent with the gauge-invariant analysis conducted in Ref. \cite{JentschuraPRL2015}.

\section{Discussion}
The theory of  the friction force at the relativistic movement of a small neutral particle relative to the  blackbody radiation was developed by Dedkov and Kyasov \cite{DedkovPLA2005,DedkovNIMPR2010,DedkovPhysScr2014} (DK) in a series of works in which the friction force in the $K$ frame was calculated. Recently DK's results were confirmed within the framework of a fully covariant theory \cite{HenkelNJP2013}. In Ref.\cite{VolokitinPRB2008}  we proposed an alternative theory (VP) in which the friction force in the $K^{\prime}$ frame was calculated. This theory was criticized  by DK in Ref.\cite{DedkovNIMPR2010}. In this paper,  a general theory which includes previous theories and establishes a link between them is developed. Within this theory we refute  the criticisms  of the  VP theory made by  DK. The key point in the VP theory  is the establishment of the relation  between the friction forces in the $K$ and $K^{\prime}$ frames. According to the VP theory  these friction forces are generally not equal but differ due to the change in the rest mass of a particle  as a result of absorption and emission of radiation  by a particle. This result is in  sharp contradiction to the opinion of DK  in  Ref.\cite{DedkovNIMPR2010} which overlooked   the  effect of changing the rest mass of a  particle. As a result, they came to the  conclusion (which we find to be incorrect) that in the general case the total friction forces, which consist from the contributions   from the interaction of a particle with the blackbody radiation and with its own  radiation, in the $K$ and $K^{\prime}$ frames are equal. DK suggested \cite{DedkovNIMPR2010} that, in the general case, the contribution to the friction force from the interaction of a particle with its own radiation  in the $K^{\prime}$ frame is not equal to zero, which we believe is incorrect. In the present paper the result that this frictional force is equal to zero is proved by rigorous  calculation. However, this result is evident from the symmetry of the dipole radiation in the $K^{\prime}$ frame. Due to this symmetry  the momentum emitted by a particle, and hence the force due to the interaction of a particle with   its own radiation, is equal to zero. So, if we follow the reasoning of DK, the total friction force in the $K$ frame in general case should be equal to the contribution to the friction force in the $K^{\prime}$ frame  due to the interaction of a particle with the blackbody radiation, which we find to be incorrect   and contradicts     opinions stated by DK. As shown in the present work, the equality of  the total friction forces  in the $K$ and $K^{\prime}$ frames takes place   only in the special (but physically  most important) case of  motion of a particle with a steady- state temperature. In this case the particle rest mass does not change and friction forces in the $K$ and $K^{\prime}$ frames are equal to the friction force  in the  $K^{\prime}$ frame  due to the interaction of a particle with the blackbody radiation, which also contradicts  the opinion of DK. In addition, considering the relativistic dynamics DK stated that the acceleration of the particle is determined by the friction force in the $K$ frame, which can be positive or negative. However, according to the  VP theory    the  particle acceleration is determined by the friction force in the  $K^{\prime}$ frame, which is always negative.

\section{Conclusion}
It was shown that in the most simple way the friction force on a neutral particle at the relativistic motion relative to the blackbody radiation can be calculated in the rest frame of a particle ($K^{\prime}$ frame) where the friction force is determined only by the interaction of a particle with the blackbody radiation. In the frame of the blackbody radiation ($K$ frame) the friction force can be found by using the Lorentz transformation for the friction force. The friction force in the $K$ frame is determined by both the interactions of a particle with the blackbody radiation and the thermal radiation of a particle and can be either positive or negative. However, the acceleration, which is determined by the friction force
in the $K^{\prime}$ frame, is always negative. The difference between the forces in the different frames is connected to a change in the rest mass of a particle due to the absorption and emission of radiation by a particle.  For the steady-state temperature of a particle the friction forces in the $K^{\prime}$ and $K$ frames are equal. For an atom, blackbody radiation friction is determined by the radiation electronic linewidth broadening, which can be calculated by considering the interaction of an atom with its own radiation. The obtained results can be used to studying  the interaction of cosmic rays with background blackbody radiation and noncontact friction. For example, the approach proposed in this paper  for calculating the radiation friction can  also be used to calculate quantum Vavilov-Cherenkov radiation \cite{KardarPRA2013} by a neutral particle moving parallel to a dielectric.

\vskip 0.5cm
The study was supported by the Ministry of education and science
 of Russia under Competitiveness Enhancement Program of SSAU
for 2013-2020 years,
 the Russian Foundation for Basic
Research (Grant N 14-02-00384-a) and COST Action MP1303 ``Understanding and Controlling Nano and Mesoscale Friction.''

\end{document}